\documentclass[aps,prc,twocolumn,superscriptaddress,nofootinbib,longbibliography,floatfix,10pt]{revtex4-1}

\usepackage{graphicx}
\usepackage{dcolumn}
\usepackage{bm}
\usepackage{morefloats}
\usepackage{multirow}
\usepackage{amssymb}
\usepackage{amsmath}
\usepackage{xcolor} 
\usepackage{longtable}
\usepackage{fix-cm}
\usepackage{mathptmx} 
\usepackage[T1]{fontenc}
\usepackage[colorlinks,allcolors=blue]{hyperref}
\makeatletter
\def\NAT@def@citea{\def\@citea{\NAT@separator}}
\makeatother

\begin{document} 

\title{Impact of the initial fluctuations on the dissipative dynamics of interacting Fermi systems: A model case study}

\author{Ibrahim Ulgen}
\affiliation{Physics Department, Faculty of Sciences, Ankara University, 06100 Ankara, Turkey}
\affiliation{Physics Department, Faculty of Sciences and Arts, Siirt University, 56100 Siirt, Turkey}
\author{Bulent Yilmaz}\email{bulent.yilmaz@science.ankara.edu.tr}
\affiliation{Physics Department, Faculty of Sciences, Ankara University, 06100 Ankara, Turkey}
\author{Denis Lacroix}
\affiliation{Institut de Physique Nucl\'eaire, IN2P3-CNRS, Universit\'e Paris-Sud, Universit\'e Paris-Saclay, F-91406 Orsay Cedex, France}

\date{\today}

\begin{abstract} 
Standard methods used for computing the dynamics of a quantum many-body system are the mean-field (MF) approximations such as the time-dependent Hartree-Fock (TDHF) approach. Even though MF approaches are quite successful, they suffer some well-known shortcomings, one of which is insufficient dissipation of collective motion. The stochastic mean-field approach (SMF), where a set of MF trajectories with random initial conditions are considered, is a good candidate to include dissipative effects beyond mean field. In this approach, the one-body density matrix elements are treated initially as a set of stochastic Gaussian c numbers that are adjusted to reproduce first and second moments of collective one-body observables. It is shown that the predictive power of the SMF approach can be further improved by relaxing the Gaussian assumption for the initial probabilities. More precisely, using Gaussian or uniform distributions for the matrix elements generally leads to overdamping for long times, whereas distributions with smaller kurtosis lead to much better reproduction of the long time evolution.
\end{abstract}


\maketitle

\section{Introduction}
\label{sec1}
In many situations, the evolution of a quantum system can be replaced by a set of classical evolutions with random initial conditions optimized to reproduce at best the initial 
quantum zero-point motion~\cite{davis1984}. This quantum-to-classical mapping is particularly suitable in the absence of interference or tunneling and can be exact in some cases, like 
the free wave expansion or the quantum harmonic oscillator. This idea is employed in many fields of physics to describe the out-of-equilibrium motion of complex systems, like in quantum optics~\cite{gardiner2004}. This also includes the possibility to describe many-body interacting systems. An illustration in
bosonic systems is the truncated Wigner approximation (TWA)~\cite{sinatra2002} (see also~\cite{polkovnikov2010}). For Fermi systems, as underlined 
in ref.~\cite{davidson2017}, such mapping is more tricky due to the absence of natural classical representation, in contrast to bosonic systems.  
Despite this inherent difficulty for fermionic systems, one can mention two attempts to map the complex many-body problem of interacting systems as a set of "classical trajectories".
The first one is the stochastic mean-field (SMF) approach~\cite{ayik2008,lacroix2014}, where the stochastic one-body density matrices are treated as classical objects evolving through the time-dependent mean-field (TDMF) equation of motion. A second formulation was  
made more recently in ref.~\cite{davidson2017} based on the mapping between fermionic and bosonic operators, leading to the 
fermionic TWA (f-TWA). We actually realized recently that the equation of motion obtained in the f-TWA often coincides with the TDMF evolution, and therefore these two approaches 
are relatively close to each other. In recent decades, the SMF approach has been successfully applied to model cases~\cite{lacroix2012,lacroix2014a,yilmaz2014a}    
as well as realistic simulations of dynamical phenomena in nuclear physics~\cite{ayik2009,ayik2010,yilmaz2011,yilmaz2014,tanimura2017,ayik2017,ayik2018,ayik2019}. It has also been extended to describe superfluid systems in ref.~\cite{lacroix2013}.  
The SMF approach and the f-TWA approach have also in common that they both assume Gaussian probabilities for the initial fluctuations. However, this assumption turns out to
be guided more by practical arguments than by first principles. In the present work, we further explore the possibility of using an alternative probability distribution for the initial conditions in the SMF approach. We show that, while the second moment of a one-body observable can be generally interpreted classically, the fourth moment of a one-body observable is more problematic, since it can lead to negative values for the fourth moments of the stochastic matrix elements of one-body densities when considering many-body Fermi systems. Such behavior cannot be reproduced by a classical mapping. However, probability distributions with smaller kurtosis compared to the Gaussian distribution turn out to be more efficient 
to describe the time evolution in the SMF approach. Such finding is illustrated on a modified version of the Lipkin-Meshkov-Glick model~\cite{lacombe2016}. 

One of the interesting aspects of SMF is that it provides a fully microscopic description of fluctuation and dissipation. It also can be used to make connection with the more phenomenological Langevin approach in collective space where few relevant collective degrees of freedom are preselected~\cite{ayik2008}. The Langevin technique has been successfully applied to dissipative nuclear process like fission and/or heavy-ion collisions~\cite{abe1996,karpov2001,zagrebaev2005,aritomo2009,wen2013,karpov2017,sekizawa2019,lin2019}. Although we are still trying to quantify and/or improve the predictive power of the SMF approach, in the long term it might provide a fully microscopic framework for dissipative process without needing assumptions inherent to more phenomenological approaches, like the selection of collective degrees of freedom or the assumption on the nature of the noise.

The paper is organized as follows. In Sec.~\ref{sec2}, the third and fourth moments of the matrix elements of the stochastic one-body densities are derived within the SMF approach and it is shown that  initial probability distribution functions with small kurtosis can provide a better approximation to the fourth central moment. In Sec.~\ref{sec3}, the SMF dynamics is applied to a modified version of the Lipkin-Meshkov-Glick model. Finally, the conclusions are given in Sec.~\ref{sec4}.
 
\section{Stochastic mean-field approach}
\label{sec2}
The Schr\"odinger equation of a many-body fermionic system can be cast into a hierarchy of differential equations for reduced densities (one-body, two-body, etc.) known as BBGKY (Bogoliubov-Born-Green-Kirkwood-Yvon) hierarchy. The truncation of BBGKY equations at lowest order gives the mean-field equation for the dynamics of one-body density,
\begin{align}\label{eq1}
i\hbar\frac{\partial}{\partial t} \rho= [h[\rho],\rho],
\end{align}
where $h[\rho]$ is the mean-field Hamiltonian. In the mean-field approximation, one-body density contains all the information on the system, hence the many-body state is restricted to be a Slater determinant during the dynamical evolution. MF dynamics (or TDHF) is known to provide good approximation to one-body observables but severely underestimate their quantum fluctuations. Beyond -mean-field approaches are necessary to overcome these shortcomings and provide a more accurate description of the dynamical evolution. 

In the SMF approach~\cite{ayik2008}, the initial one-body density $\rho(0)$ is replaced by an ensemble of stochastic initial one-body densities $\rho^\lambda(0)$, where $\lambda$ stands for event label. Each of these densities evolves with its self-consistent mean-field equation,  
\begin{align}\label{eq2}
i\hbar\frac{\partial}{\partial t} \rho^\lambda= [h[\rho^\lambda],\rho^\lambda].
\end{align}
For each event $\lambda$, the ''event`` expectation value of a one-body observable $A$ is given by
\begin{align}\label{ola}
\langle A\rangle^\lambda= {\rm Tr}({\rho}^\lambda A),
\end{align}  
\begin{flushright}
where ''event`` means one trajectory associated with a specific distribution for the stochastic initial one-body densities.
\end{flushright}
In the SMF approach, the expectation values are obtained by ensemble averages. Hence, the expectation value of the one-body observable is defined as,
\begin{align}\label{smf_mean}
\overline{\langle A\rangle^\lambda}=&{\rm Tr}(\overline{{\rho}^\lambda} A)\nonumber\\
=&\sum_{ij}\overline{\rho_{ij}^\lambda} A_{ji},
\end{align}
where the overline stands for ensemble averaging explicitly given by
\begin{align}
\overline{\rho_{ij}^\lambda}=\lim_{\mathcal{N}\rightarrow \infty}\frac{1}{\mathcal{N}}\sum_{\lambda=1}^{\mathcal{N}}\rho_{ij}^\lambda.
\end{align}
Here, $\mathcal{N}$ is the number of events in the ensemble. The quantum variance of the one-body operator is defined by  
\begin{align}\label{smf_var}
\sigma^{(2)}_A=&\overline{\left(\langle A\rangle^\lambda
-\overline{\langle A\rangle^\lambda}\right)^2}\nonumber\\
=&\overline{\left[{\rm Tr}(\delta{\rho}^\lambda\, A)\right]^2}\nonumber\\
=&\sum_{ijkl}\overline{\delta\rho^\lambda_{ij}\delta\rho^\lambda_{kl}}A_{ji}A_{lk},
\end{align}
where $\delta\rho^\lambda= \rho^\lambda -\overline{\rho^\lambda}$. 

In the SMF approach, the distribution of the stochastic matrix elements, $\rho_{ij}^\lambda(0)$, is chosen such that the initial mean and variances of the observables are the same with those obtained with the initial density $\rho(0)$. If $\{|i\rangle\}$ stands for the natural basis which satisfies $\langle i|\rho(0)|j\rangle=n_i\delta_{ij}$, the mean and variance of $A$ are the expectation values given by
\begin{align}\label{q_mean}
\langle A(0)\rangle=\sum_i n_i A_{ii},
\end{align}
and
\begin{align}\label{q_var}
\langle A^2(0)\rangle-\langle A(0)\rangle^2=\frac{1}{2}\sum_{ij}\left[n_i(1-n_j)+n_j(1-n_i)\right] A_{ji}A_{ij},
\end{align}
respectively. Comparing Eqs.~(\ref{smf_mean}) and (\ref{smf_var}) with Eqs.~(\ref{q_mean}) and (\ref{q_var}), we see that the quantum averages match the ensemble averages if we have,
\begin{align}\label{mo1}
\overline{\rho^\lambda_{ij}(0)}=&n_i\delta_{ij},\\
\label{mo2}
\overline{\delta\rho^\lambda_{ij}(0)\delta\rho^\lambda_{kl}(0)}=&\delta_{il}\delta_{jk}\frac{1}{2}\left[n_i(1-n_j)+n_j(1-n_i)\right].
\end{align}

In the original formulation of the SMF approach~\cite{ayik2008}, the stochastic matrix elements $\delta\rho^\lambda_{ij}(0)$ are assumed to be Gaussian random numbers satisfying Eqs.~(\ref{mo1}) and (\ref{mo2}). 

\subsection*{Higher order moments of the stochastic matrix elements of the one-body density}
Here, we derive higher order moments for the stochastic matrix elements and test the Gaussian assumption of the original formulation of the SMF approach when the initial state is a Slater determinant. For this purpose, similarly to the second central moment Eq.~(\ref{smf_var}), we define the expectation value of the mth central moment of a one-body operator $A$ as
\begin{align}
\sigma^{(m)}_A=&\overline{\left(\langle A\rangle_\lambda
-\overline{\langle A\rangle_\lambda}\right)^m}\nonumber\\
=&\overline{\left[{\rm Tr}(\delta{\rho}^\lambda\, A)\right]^m},
\end{align}
where $\delta\rho^\lambda= \rho^\lambda -\overline{\rho^\lambda}$. Hence, the third and fourth central moments can be written as,
\begin{align}\label{smf3}
\sigma^{(3)}_A=&\sum_{ijklmn}\overline{\delta\rho^\lambda_{ij}\delta\rho^\lambda_{kl}\delta\rho^\lambda_{mn}}A_{ji}A_{lk}A_{nm},\\
\label{smf4}
\sigma^{(4)}_A=&\sum_{ijklmnpr}\overline{\delta\rho^\lambda_{ij}\delta\rho^\lambda_{kl}\delta\rho^\lambda_{mn}\delta\rho^\lambda_{pr}}A_{ji}A_{lk}A_{nm}A_{rp}.
\end{align}
The corresponding quantum central moments, for an initial Slater determinant, are given by (see the Appendix for details)
\begin{align}\label{qm3}
\langle(\Delta A)^3\rangle=&\sum_{ijk} \Lambda^{(3)}_{ijk}\,A_{ij}A_{jk}A_{ki},\\
\label{qm4}
\langle(\Delta A)^4\rangle=&\sum_{ijkl} \Lambda^{(4a)}_{ijkl}\,A_{ij}A_{jk}A_{kl}A_{li}\nonumber\\
&\!+3\sum_{ijkl} \Lambda^{(4b)}_{ijkl}\,A_{ij}A_{ji}A_{kl}A_{lk},
\end{align}
where $\Delta A= A-\langle A\rangle$ and
\begin{align}
\label{p3}
\Lambda^{(3)}_{ijk}=&\; \frac{1}{3}\left[n_i(1-3n_j)(1-n_jn_k)+n_k(1-3n_i)(1-n_in_j)\right.\nonumber\\
&\quad \left. + n_j(1-3n_k)(1-n_kn_i)\right],\\
\label{p4a}
\Lambda^{(4a)}_{ijkl}=&\;\frac{1}{4}\left[ n_i(1-4n_j)(1-3n_k)(1-n_jn_kn_l)\right.\nonumber\\
&\quad+n_l(1-4n_i)(1-3n_j)(1-n_in_jn_k)\nonumber\\
&\quad+n_k(1-4n_l)(1-3n_i)(1-n_ln_in_j)\nonumber\\
&\quad\left.+n_j(1-4n_k)(1-3n_l)(1-n_kn_ln_i)\right],\\
\label{p4b}
\Lambda^{(4b)}_{ijkl} =&\;\frac{1}{8}\left\{ n_in_k\left[(1-2n_l)(1-n_ln_j)+(1-2n_j)(1-n_jn_l)\right]\right.\nonumber\\
&\quad+n_jn_k\left[(1-2n_l)(1-n_ln_i)+(1-2n_i)(1-n_in_l)\right]\nonumber\\
&\quad+n_in_l\left[(1-2n_k)(1-n_kn_j)+(1-2n_j)(1-n_jn_k)\right]\nonumber\\
&\quad\left.+n_jn_l\left[(1-2n_k)(1-n_kn_i)+(1-2n_i)(1-n_in_k)\right]\right\}.
\end{align}
Comparing Eqs.~(\ref{smf3}) and (\ref{smf4}) with Eqs.~(\ref{qm3}) and (\ref{qm4}), we see that the quantum and ensemble averages coincide if we have
\begin{align}
\label{mom3}
\overline{\delta\rho^\lambda_{ij}(0)\delta\rho^\lambda_{kl}(0)\delta\rho^\lambda_{mn}(0)}=&\delta_{il}\delta_{jm}\delta_{kn}\Lambda^{(3)}_{jik},\\
\label{mom4}
\overline{\delta\rho^\lambda_{ij}(0)\delta\rho^\lambda_{kl}(0)\delta\rho^\lambda_{mn}(0)\delta\rho^\lambda_{pr}(0)}
=&\delta_{il}\delta_{kn}\delta_{rm}\delta_{jp}
\Lambda^{(4a)}_{jikr}\nonumber\\
&+\delta_{il}\delta_{jk}\delta_{rm}\delta_{pn}
3\Lambda^{(4b)}_{jipr}.
\end{align}

By considering that all initial densities $\rho^\lambda(0)$ in the ensemble are Hermitian [in the following, we drop the labels for brevity, i.e.,~$\delta\rho_{ij}\equiv \delta\rho_{ij}^\lambda(0)$.], it immediately follows from Eq.~(\ref{mo2}) that $\overline{(\delta\rho_{ii})^2}= 0$ which means that 
\begin{align}\label{zero}
 \delta\rho_{ii}=0
\end{align}
and hence the matrix elements $\rho_{ii}$ are not fluctuating. Furthermore, for $i\neq j$, we have
\begin{align}\label{secmom}
\overline{r_{ij}^2} + \overline{s_{ij}^2}= 
\left\{\begin{array}{cc}
\frac{1}{2}\quad & \qquad\text{if}~(i,j)= (p,h),\\ \\
0\quad & \text{otherwise}, 
\end{array}\right.
\end{align}
where we introduced the real and imaginary parts of the matrix element dispersions, $\delta\rho_{ij}=r_{ij}+i\,s_{ij}$. Here, $(i,j)= (p,h)$ means that one of the states is a particle state and the other a hole state. The variance cannot be negative and hence, from Eq.~(\ref{secmom}), we have $\overline{r_{ij}^2}=\overline{s_{ij}^2}=0$ for $(i,j)= (p,p)$ or $(i,j)=(h,h)$ which means that these matrix elements are zero. Note that the stochastic matrix elements satisfy the equations $r_{ji}=r_{ij}$ and $s_{ji}=-s_{ij}$ due to Hermiticity.

In order to deduce the properties required for $\delta\rho$ to fulfill the moments Eqs.~(\ref{mom3}) and (\ref{mom4}), we need to analyze the possible values of the $\Lambda$ terms.  In general, from Eqs.~(\ref{p3})-(\ref{p4b}), the $\Lambda$ terms can assume the following nonzero values:
\begin{align}\label{lam}
\Lambda^{(3)}_{pph}&=\Lambda^{(3)}_{php}=\Lambda^{(3)}_{hpp}=-\frac{1}{3},\nonumber\\
\Lambda^{(3)}_{phh}&=\Lambda^{(3)}_{hph}=\Lambda^{(3)}_{hhp}=+\frac{1}{3},\nonumber\\
\Lambda^{(4a)}_{ppph}&=\Lambda^{(4a)}_{pphp}=\Lambda^{(4a)}_{phpp}=\Lambda^{(4a)}_{hppp}=+\frac{1}{4},\nonumber\\
\Lambda^{(4a)}_{hhhp}&=\Lambda^{(4a)}_{hhph}=\Lambda^{(4a)}_{hphh}=\Lambda^{(4a)}_{phhh}=+\frac{1}{4},\nonumber\\
\Lambda^{(4a)}_{pphh}&=\Lambda^{(4a)}_{hhpp}=\Lambda^{(4a)}_{hpph}=\Lambda^{(4a)}_{phhp}=-\frac{1}{2},\nonumber\\
\Lambda^{(4a)}_{phph}&=\Lambda^{(4a)}_{hphp}=-1,\nonumber\\
\Lambda^{(4b)}_{phph}&=\Lambda^{(4b)}_{hphp}=\Lambda^{(4b)}_{hpph}=\Lambda^{(4b)}_{phhp}=+\frac{1}{4},
\end{align}
where p stands for a particle and h stands for a hole state. The nonzero terms of Eq.~(\ref{mom3}) are given by
\begin{align}
\label{mom3a}
\overline{\delta\rho_{ij}\delta\rho_{ki}\delta\rho_{jk}}=&\Lambda^{(3)}_{jik}.
\end{align}
Similarly, the nonzero terms of the first and second terms of Eq.~(\ref{mom4}), by taking into account the Kronecker delta functions, read as
\begin{align}
\label{mom4a}
\overline{\delta\rho_{ij}\delta\rho_{ki}\delta\rho_{rk}\delta\rho_{jr}}=\Lambda^{(4a)}_{jikr}
\end{align}
and
\begin{align}
\label{mom4b}
\overline{\delta\rho_{ij}\delta\rho_{ji}\delta\rho_{rp}\delta\rho_{pr}}=3\Lambda^{(4b)}_{jipr},
\end{align}
respectively. 

The form of the last three equations suggests that these equations can be satisfied if one considers that the stochastic matrix elements are correlated with each other. However, averages that contain terms with the same indices $\delta\rho_{jj}$ in Eqs.~(\ref{mom3a}) and (\ref{mom4a}) lead to, for instance, $\overline{\delta\rho_{ij}\delta\rho_{ji}\delta\rho_{jj}}=\Lambda^{(3)}_{jij}$ and $\overline{\delta\rho_{ij}\delta\rho_{ii}\delta\rho_{ji}\delta\rho_{jj}}=\Lambda^{(4a)}_{jiij}$,
where left-hand side of these equations is zero due to Eq.~(\ref{zero}) and the right-hand side of the equations is non-zero as seen from Eq.~(\ref{lam}). Hence, Eqs.~(\ref{mom3a}) and (\ref{mom4a}) cannot be satisfied completely even with correlated matrix elements. The reason for such a discrepancy is related to the fact that probability distributions cannot simulate quantum mechanical systems. In conclusion, probability distributions can only approximately match higher quantum mechanical moments. Here, we explore probability distributions that can at best approximate the third and fourth quantum mechanical moments.    

In this study, we investigate the consequences of the simple case of uncorrelated matrix elements. Hence, we assume that each stochastic matrix element is statistically independent of the others as well as that the real and imaginary parts of a matrix element are also statistically independent. Then, for simplicity, we set $\overline{\delta\rho_{ij}\delta\rho_{ji}\delta\rho_{ij}}=\overline{r_{ij}^3} + i\overline{s_{ij}^3}=0$ for the third central moments, which means 
\begin{align}
\overline{r_{ij}^3} = \overline{s_{ij}^3}=0.
\end{align}
This is in accordance with the original formulation of the SMF approach since the third central moment of a mean-zero Gaussian random number is zero. Note that Eq.~(\ref{mom3a}) does not provide any restrictions for the $\overline{\delta\rho_{ij}\delta\rho_{ji}\delta\rho_{ij}}$ moments. However, a condition for the fourth central moments can be obtained by assuming uncorrelated matrix elements. It is seen from Eqs.~(\ref{mom4a}) and (\ref{mom4b}) that one has
\begin{align}
\overline{\delta\rho_{ij}\delta\rho_{ji}\delta\rho_{ij}\delta\rho_{ji}}=\Lambda^{(4a)}_{jiji}+3\Lambda^{(4b)}_{jiji},
\end{align}
which leads to the result,
\begin{align}\label{kur}
\overline{r_{ij}^4} + \overline{s_{ij}^4} +
2\,\overline{r_{ij}^2}\;\overline{s_{ij}^2}= 
\left\{\begin{array}{cc}
-\frac{1}{4}\quad & \qquad \text{if}~ (i,j)= (p,h),\\ \\
\text{ }\;\;0\quad & \text{otherwise},
\end{array}\right.
\end{align}
where Eq.~(\ref{lam}) was used.

No probability distribution can have a negative second and/or fourth central moment. Hence, the left-hand side of Eq.~(\ref{kur}) can never be negative. However, the distributions that we want to obtain are quantum mechanical in nature. Actually the noncommutativity of operators within quantum mechanics is the main reason for inadequacy of probability distributions to simulate the quantum systems. 
In the phase-space formulation of quantum mechanics, we already know such behaviors coming from the Wigner distribution which can assume negative values and therefore is called a quasiprobability distribution~\cite{bednorz2011}. It is important to note that positive-definite phase-space distributions such as the Husimi distribution exist; however, they are still quasiprobability distributions since the averages are taken by the Weyl symbol of the operators rather than the operators themselves~\cite{yilmaz2014a}. 

Based on the discussion above, since a probability distribution cannot lead to a negative fourth central moment, we anticipate that an efficient approximation would be to consider a distribution which minimizes the left-hand side of Eq.~(\ref{kur}) while satisfying Eq.~(\ref{secmom}). Equation~(\ref{secmom}) sets a condition on the sum of the variances and hence it does not impose any conditions on the weights to the variances of the real and imaginary parts. In order to investigate the dependence to the weights, we define a parameter $\chi=\overline{r^2_{ij}}$. From Eq.~(\ref{secmom}), we have $\overline{s^2_{ij}}=1/2 - \chi$, where $0\le \chi\le 1/2$. Substituting these results into the left-hand side of Eq.~(\ref{kur}) we obtain the function
\begin{align}\label{func}
F(\chi,\gamma)= 2(\gamma - 1)\chi^2 -(\gamma -1)\chi+\frac{\gamma}{4},  
\end{align}
where we used the kurtosis of the distribution, which is defined as 
\begin{align}\label{kurdef}
\gamma= \frac{\overline{r^4_{ij}}}{\left(\overline{r^2_{ij}}\right)^2}.
\end{align}
\begin{figure}[!hpt]
\vspace{0.2cm}
\includegraphics*[width=8.5cm]{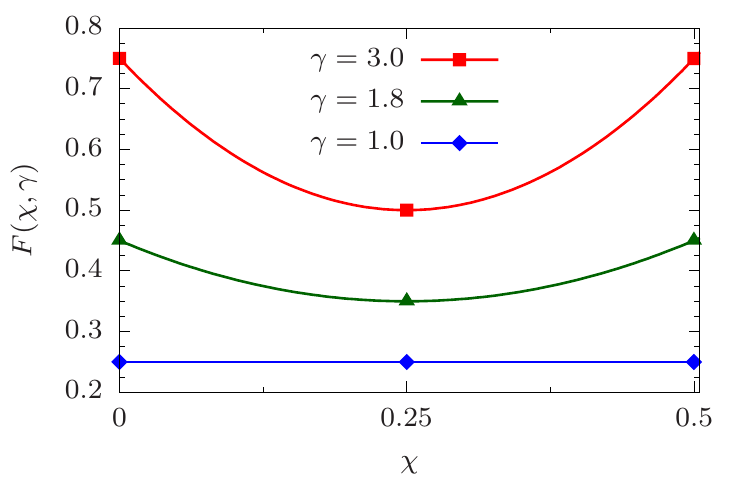}
\vspace{0.2cm}
\caption{The plot of the function (\ref{func}) versus the parameter $\chi$ for three different kurtoses $\gamma$: $\gamma=3$, $\gamma=1.8$, and $\gamma=1$ correspond to the Gaussian, uniform, and two-point distributions, respectively.}
\label{fig1}
\end{figure}
The kurtosis is an invariant form of the fourth central moment that only depends on the distribution and does not depend on the particular values of the variance. Figure~\ref{fig1} shows the plot of the function (\ref{func}) versus the weight parameter $\chi$ for the kurtosis values $3$ and $1.8$, which are the kurtoses of the Gaussian and the uniform distributions, respectively. The kurtosis value of $1$ corresponds to the minimum possible value. The probability distribution function with the minimum kurtosis is the so-called two-point distribution function which can take only two values, $\sigma$ and $-\sigma$, with equal probabilities~\cite{moors1988}. This probability function can be written as
\begin{align}\label{twop}
P_{2p}(x)= \frac{1}{2}\delta(x+\sigma)+\frac{1}{2}\delta(x-\sigma),
\end{align}
where $\delta(x)$ is the Dirac delta function and $\sigma$ is the standard deviation of $x$. It is seen from Fig.~\ref{fig1} that the function $F$ decreases as the kurtosis $\gamma$ decreases. We show below that the SMF dynamics with a small kurtosis probability distribution provides a better approximation to the exact dynamics than the SMF dynamics with a large kurtosis distribution. In particular, for the three distributions we considered here, the two-point distribution provides the best approximation and the Gaussian distribution provides the worst approximation. The choice of the weight parameter $\chi$ also has an impact on the values of the function $F$, as seen from Fig.~\ref{fig1}. The equal weight case, $\chi=\overline{r^2_{ij}}=\overline{s^2_{ij}}=1/4$, provides the minimal value of $F$ for a fixed value of the kurtosis $\gamma$ except for the two-point distribution which is constant. Hence, the case with equal weights for the variances of the real and imaginary parts of the stochastic matrix elements is anticipated to provide a better approximation than any unequal weights case.

\section{Applications}
\label{sec3}
In this section, we apply the SMF approach with the two-point distribution as well as the Gaussian and uniform distributions on an exactly solvable model and show that indeed the SMF dynamics with the two-point distribution for the matrix elements gives a better agreement with the exact dynamics than the SMF dynamics with the uniform or Gaussian distributions.

\subsection{A modified Lipkin-Meshkov-Glick model}
\begin{figure}[!hpt]
\vspace{0.2cm}
\includegraphics*[width=8.5cm]{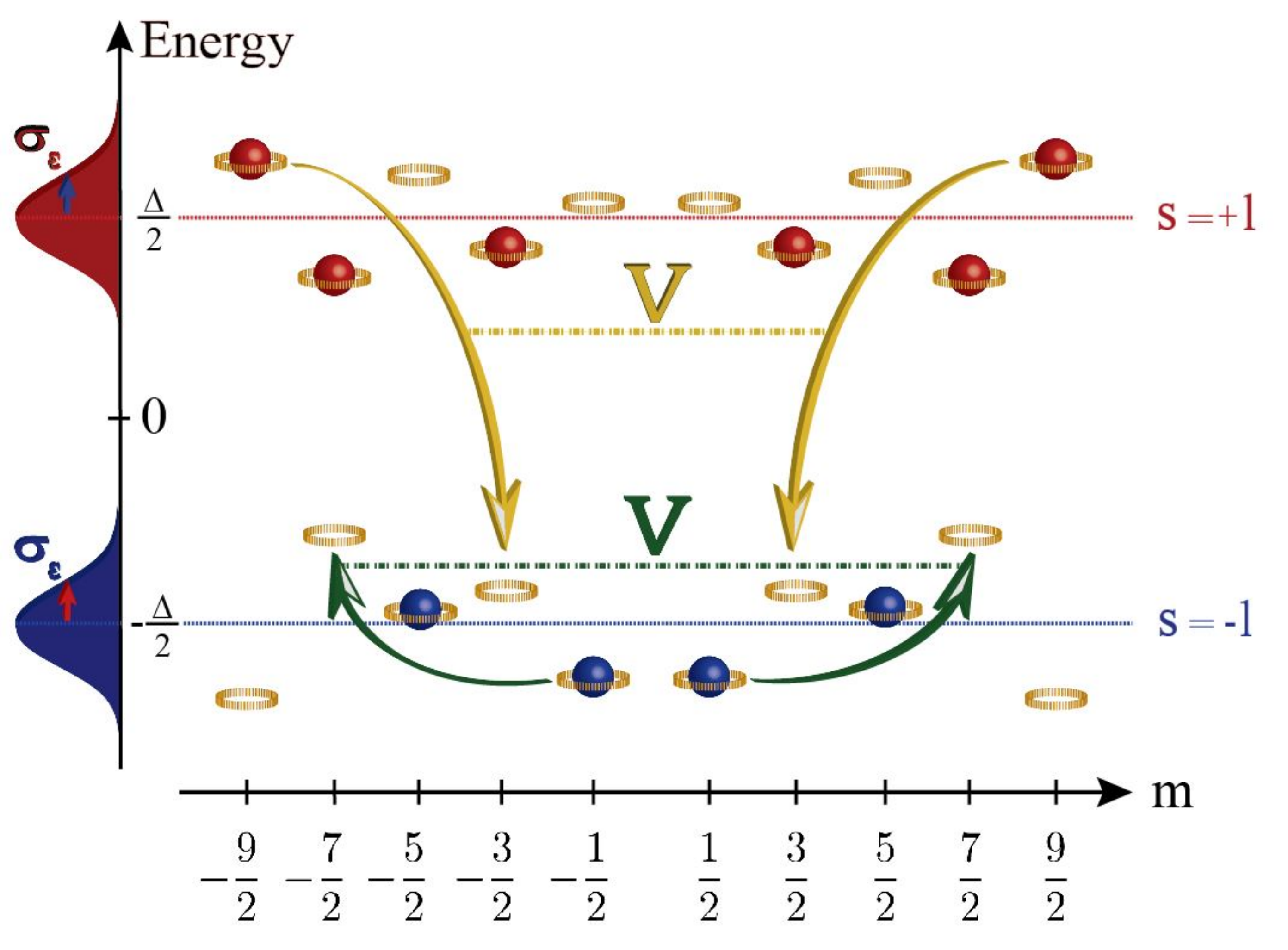}
\vspace{0.2cm}
\caption{Schematic illustration of the mLMG model. Two transitions are indicated: one within the same level (green arrows) and another between the two levels (yellow arrows).}
\label{fig2}
\end{figure}

We consider a modified Lipkin-Meshkov-Glick model (mLMG) that was introduced in ref.~\cite{lacombe2016}. It has a pairing Hamiltonian structure and the single particle energies are stochastically distributed over the levels allowing for a description of dissipation. The Hamiltonian of this model reads 
\begin{eqnarray}\label{ham1}
H = T + V
\end{eqnarray}
with 
\begin{eqnarray}
T &=& \sum_\alpha \frac{s_\alpha \varepsilon_\alpha}{2} a^\dagger_\alpha a_\alpha , \\
V &=& v_0 S_+ S_-, 
\end{eqnarray}
where $\alpha=(s_\alpha,m_\alpha)$, $s_\alpha \in \{-1,+1\}$, $m_\alpha \in \{-j,-j+1,...,j-1,j\}$, $S_+ = \sum_{\alpha>0} a^\dagger_\alpha a^\dagger_{\underline{\alpha}}$, and $S_-=S_+^\dag$. Here, $\underline{\alpha}=(s_\alpha,-m_\alpha)$ and $\alpha>0$ means $m_\alpha>0$. $v_0$ is the interaction strength and $\varepsilon_\alpha/2$ are the single-particle energies which are random numbers with some variance $\sigma^2_\varepsilon$. The mean values of the single-particle energies in the upper level $s_\alpha=1$ and in the lower level $s_\alpha=-1$ are $\Delta/2$ and $-\Delta/2$, respectively. In order to avoid any ambiguity, we should state that the stochastic single-particle energies are only set once and these values are used for the dynamics. Figure~\ref{fig2} shows a schematic illustration of the mLMG model. Two transitions are indicated on the figure. Note that, contrary to the original LMG model~\cite{lipkin1965}, there are transitions within the same level, which is due to the fact that the single-particle energies are different.  

\subsection{Exact, mean-field, and stochastic mean-field dynamics}
The exact dynamics, for an initial state $|\Psi(0)\rangle$, is formally given by
\begin{align}
 |\Psi(t)\rangle= e^{-iHt/\hbar}|\Psi(0)\rangle.
\end{align}
In practice, there are several numerical methods to solve this equation of motion. We use the iterated Crank-Nicolson method for computing the exact dynamics~\cite{teukolsky2000}. 

The mean-field equation of motion (or TDHF equation) can be derived by using the Ehrenfest theorem,
\begin{align}
i \hbar  \frac{d\rho_{\alpha\beta}}{dt}&=i \hbar \frac{d \langle a^\dagger_{\beta}\, a_\alpha^{} \rangle }{dt}\nonumber\\
&= \langle [ a^\dagger_{\beta}\, a_\alpha^{}  , H ]\rangle,
\end{align}
where the expectation value is taken with respect to a Slater determinant. By using Eq.~(\ref{ham1}), the last equation gives
\begin{align}\label{lmgmf}
i \hbar  \frac{d\rho_{\alpha\beta}}{dt}
=&\frac{1}{2}\left(s_\alpha\varepsilon_\alpha-s_\beta\varepsilon_\beta\right)\rho_{\alpha\beta}\nonumber\\
&+v_0\sum_{\gamma>0}\left(\rho_{\underline{\gamma}\underline{\alpha}}\rho_{\gamma\beta}-\rho_{\alpha\gamma}\rho_{\underline{\beta}\underline{\gamma}}\right)\nonumber\\
&-v_0\sum_{\gamma>0}\left(\rho_{\gamma\underline{\alpha}}\rho_{\underline{\gamma}\beta}-\rho_{\alpha\underline{\gamma}}\rho_{\underline{\beta}\gamma}\right).
\end{align}
The SMF equation is directly obtained from Eq.~(\ref{lmgmf}) by replacing all one-body densities with the stochastic ones, $\rho \rightarrow \rho^\lambda$.

There are two different single-particle bases that are used in this study. One of the bases is the fixed basis of the model, which consists of the single-particle states associated with the creation and annihilation operators $a^\dagger_{\alpha}$, $a_{\alpha}$ appearing in the Hamiltonian Eq.~(\ref{ham1}). We use the labels $\alpha$, $\beta$, etc. for the fixed basis. The other basis is the natural basis which is obtained by diagonalizing the initial one-body density and we use the labels $i$, $j$, etc. for the states in this basis. The fixed basis do not change over time, whereas the natural basis change over time. Note that the MF equation (\ref{lmgmf}) is written in the fixed basis, whereas the properties of the matrix elements of the initial stochastic one-body densities in the SMF approach are derived in the natural basis in Sec.~\ref{sec2}. We solve the SMF equations of motion in the fixed basis. Hence, for each event we sample an initial stochastic one-body density in the natural basis, we write it in the fixed basis, and then evolve it with the SMF equation which is just the MF equation.

\subsection{The initial state}
\label{subsec2}
We consider the half-filled case for which there are $N$ particles for $2N$ single-particle states with initial states of the form
\begin{align}\label{ini}
|\Psi(0)\rangle= e^{i\mu D}|\Phi^{(\eta)}\rangle,
\end{align}
where $\mu$ is a real parameter and $|\Phi^{(\eta)} \rangle$ is a Slater determinant given by
\begin{align}
|\Phi^{(\eta)} \rangle = (a^\dagger_{\alpha_1})^{n_{\alpha_1}^{(\eta)}} (a^\dagger_{\alpha_2})^{n_{\alpha_2}^{(\eta)}} \ ... \ (a^\dagger_{\alpha_{2N}})^{n_{\alpha_{2N}}^{(\eta)}}|0 \rangle.
\end{align}
Here, $n_{\alpha}^{(\eta)}$ are occupation numbers with values 0 or 1 corresponding to the single-particle states indicated in Fig.~\ref{fig2}. The dipole operator $D$ reads
\begin{align}\label{dipol}
D=\sum_{m_\alpha} (a^\dagger_{+1,m_\alpha} a_{-1,m_\alpha}+a^\dagger_{-1,m_\alpha} a_{+1,m_\alpha}).
\end{align}
Since $D$ is a one-body operator the initial state $|\Psi(0)\rangle$ is also a Slater determinant due to the Thouless theorem~\cite{thouless1960}. The effect of the exponential term in Eq.~(\ref{ini}) is to introduce some excitation by an instant dipole boost~\cite{lacombe2016}.

We consider two different initial states for $N=6$ particles. The first state corresponds to the ground state, $|\Phi^{(0)}\rangle$, with all the particles in the lower level ($s_\alpha=-1$) excited by the dipole boost in Eq.~(\ref{ini}) with $\mu=0.8$. The second state is found by applying again the dipole boost with $\mu=0.8$ to the state $|\Phi^{(\eta)}\rangle$ for which 
\begin{align}
(s_\alpha,m_\alpha)=\left\{\left(+1,\,\pm\frac{1}{2}\right),\,\left(-1,\,\pm\frac{1}{2}\right),\,\left(+1,\,\pm\frac{3}{2}\right)\right\}
\end{align}
single-particle states are occupied (see Fig.~\ref{fig2}). The first and second states are denoted by $|\Psi(0)\rangle_1$ and $|\Psi(0)\rangle_2$, respectively.

Note that the schematic model in Fig.~\ref{fig2} is shown for 10 particles in 20 states. Here, we consider 6 particles in 12 states, therefore $m_\alpha$ can take the values $\{-5/2,-3/2,-1/2,1/2,3/2,5/2\}$.

\subsection{Results}
In the following computations, the units of energy will be written in terms of the level spacing $\Delta$ and the time units in terms of $\Delta^{-1}$. The single-particle energies are chosen as
\begin{align}
\varepsilon_{(+1,\pm 5/2)}/2=&0.225\,\Delta,\quad\varepsilon_{(-1,\pm 5/2)}/2=-0.222\Delta,\nonumber\\
\varepsilon_{(+1,\pm 3/2)}/2=&0.697\,\Delta,\quad\varepsilon_{(-1,\pm 3/2)}/2=-0.593\,\Delta,\nonumber\\
\varepsilon_{(+1,\pm 1/2)}/2=&0.578\,\Delta,\quad\varepsilon_{(-1,\pm 1/2)}/2=-0.685\,\Delta.
\end{align}
The mean of the single-particle energies in the upper level $\varepsilon_{(+1,m_\alpha)}/2$ and in the lower level $\varepsilon_{(-1,m_\alpha)}/2$ are $\Delta/2$ and $-\Delta/2$, respectively. The standard deviation of the $\varepsilon_\alpha/2$ values in each level is $\sigma_\varepsilon=0.2\,\Delta$. All the SMF results are obtained by considering $\mathcal{N}=10^6$ events.

We follow the dynamics of the dipole operator $D$ and the fermionic one-body entropy given by
\begin{align}\label{ent}
 S=-\text{Tr}\left[\rho\ln\rho + (1-\rho)\ln(1-\rho)\right],
\end{align}
where $\rho$ is the one-body density operator. The one-body entropy is a measure of departure from an uncorrelated (Slater) state and a measure of thermalization~\cite{lacombe2016}. For a Slater state, the entropy gets the value $S=0$. Hence, during the MF dynamics the entropy remains zero. The maximum value of the entropy for the half-filled case is obtained when all single particle states are half-filled, leading to the value
\begin{align}
 S=2N\ln 2.
\end{align}
In the SMF approach, the one-body entropy is computed by the expression
\begin{align}
 S=-\text{Tr}\left[\overline{\rho^\lambda}\ln\overline{\rho^\lambda} + (1-\overline{\rho^\lambda})\ln(1-\overline{\rho^\lambda})\right].
\end{align}

\begin{figure}[!t]
\vspace{0.2cm}
\includegraphics*[width=8.5cm]{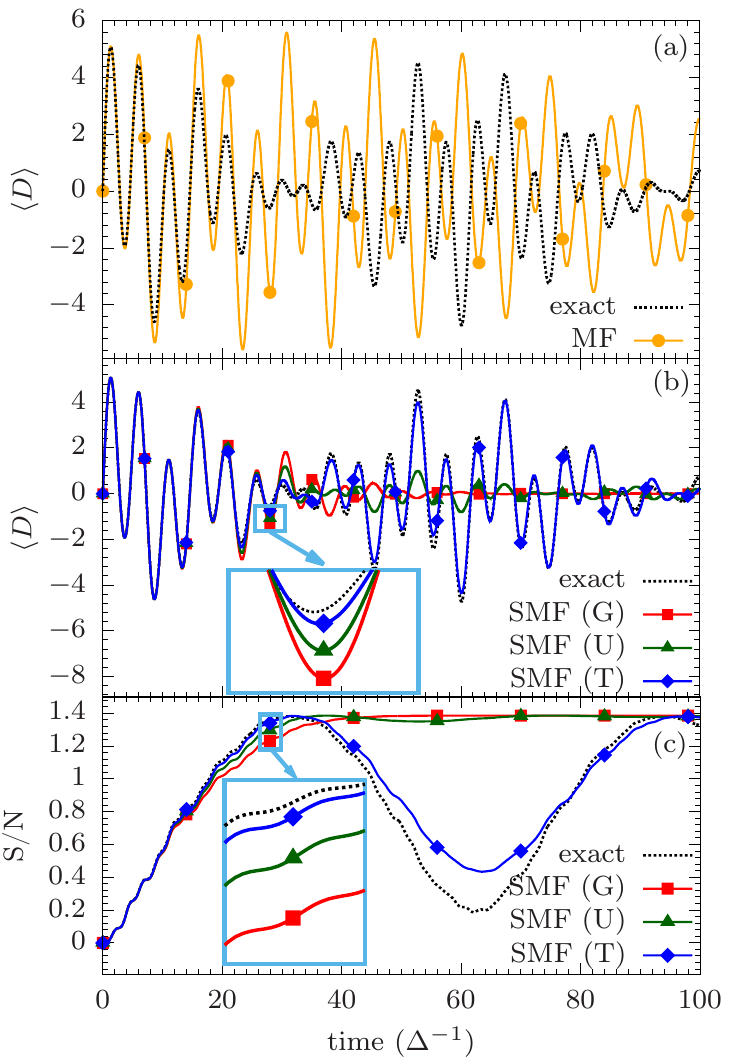}
\vspace{0.2cm}
\caption{Comparison of the dynamical evolutions of the expectation value of the dipole operator $D$ is illustrated for the exact (black line) and MF solutions (orange line with circles) (a), and for the exact and SMF solutions with Gaussian (G) (red line with boxes), uniform (U) (green line with triangles), and two-point (T) distributions (blue line with diamonds) (b). The dynamics of the one-body entropy per particle $S/N$ for the exact and SMF solutions with the same three distributions is indicated in the lower panel (c). The interaction strength is $v_0=0.05\,\Delta$ and the initial state is $|\Psi(0)\rangle_1$.}
\label{fig3}
\end{figure}
Figure~\ref{fig3} shows the expectation value of the dipole operator $D$, given by Eq.~(\ref{dipol}), and the one-body entropy $S$, given by Eq.~(\ref{ent}), per particle versus time for the weak coupling $v_0=0.05\,\Delta$. The initial state is $|\Psi(0)\rangle_1$ which is explained in the Sec.~\ref{subsec2}. In Fig.~\ref{fig3}(a), it is observed that MF evolution starts to deviate from the exact one at around $t=6\,\Delta^{-1}$, whereas in Fig.~\ref{fig3}(b) the SMF evolutions start to deviate later at around $t=20\,\Delta^{-1}$. Comparing the three SMF dynamics with the exact one in Fig.~\ref{fig3}(b), we see that the SMF approach with the two-point distribution is much better than the other two distributions at all times. A similar behavior is seen for the one-body entropy $S$ in Fig.~\ref{fig3}(c). The SMF evolution of the entropy with the two-point distribution follows the exact solution very closely, whereas SMF solutions with the Gaussian and uniform initial distributions are underestimating the exact entropy until $t=30\,\Delta^{-1}$. After that time, the exact entropy decreases, which is followed by the SMF solution with two-point distribution, whereas the solutions with the Gaussian and uniform distributions continue to increase and approach the maximum value $2\ln 2=1.38$. The solution with the uniform distribution has a small decrease from $t=40\,\Delta^{-1}$ to $t=70\,\Delta^{-1}$, hence it is slightly better than the solution with the Gaussian distribution. Note that as the kurtosis, Eq.~(\ref{kurdef}), of the distribution decreases the SMF dynamics becomes better. The best approximation is attained for the two-point distribution, Eq.~(\ref{twop}), with the minimum kurtosis value of 1. We clearly see in Fig.~\ref{fig3} that the use of the two-point distribution strongly increases the timescale over which the SMF approach is predictive.
\begin{figure}[!t]
\vspace{0.2cm}
\includegraphics*[width=8.5cm]{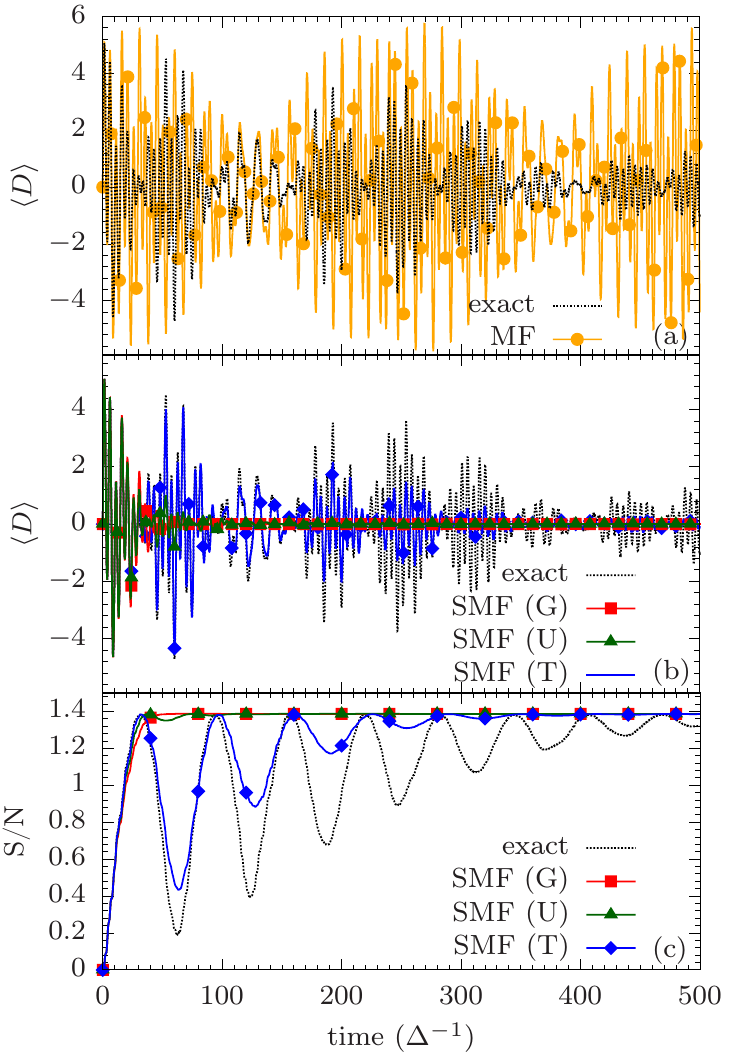}
\vspace{0.2cm}
\caption{Same as Fig.~\ref{fig3} for long time evolution}
\label{fig4}
\end{figure}

Figure~\ref{fig4} is the same as Fig.~\ref{fig3} except that the long time dynamics is shown. In Fig.~\ref{fig4}(a), the exact solution exhibits oscillations with decreasing amplitude for long times; on the other hand the MF solution has beating-like oscillations with almost constant amplitude. This result and the fact that the one-body entropy in MF approximation remains zero at all times clearly show the well-known shortcoming of MF dynamics,    
that is the underestimation of dissipation and hence thermalization in quantum many-body systems. In Fig.~\ref{fig4}(b), the SMF solution with two-point distribution follows the amplitudes of the oscillation of the exact dynamics for a longer time than the SMF solutions with Gaussian and uniform distributions, which become almost completely damped around $t=50\,\Delta^{-1}$. Figure~\ref{fig4}(c) shows the similar behavior for the entropy $S$. The SMF solutions with Gaussian and uniform distributions attain the maximum entropy value around $t=50\,\Delta^{-1}$ whereas the SMF solution with two-point distribution is in better agreement with the exact entropy for longer times.
\begin{figure}[!t]
\vspace{0.2cm}
\includegraphics*[width=8.5cm]{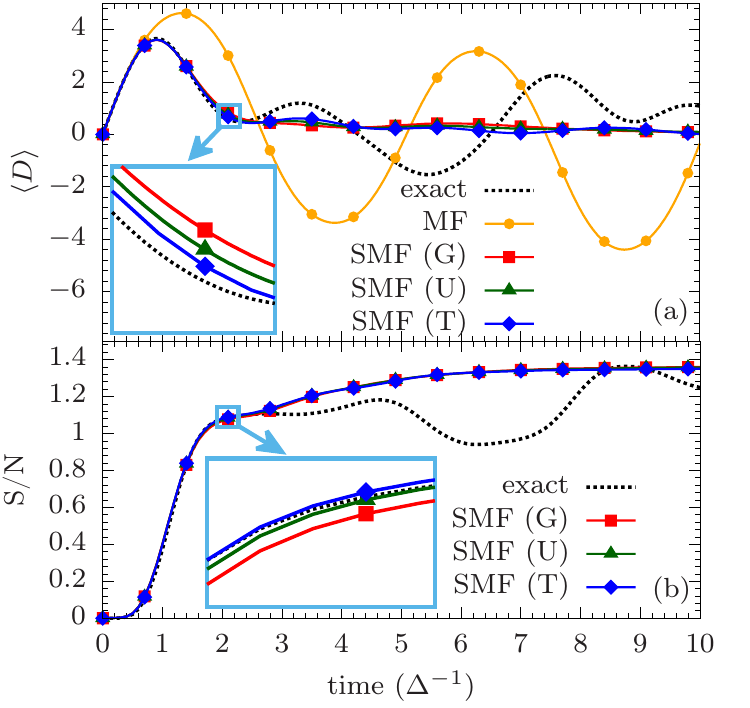}
\vspace{0.2cm}
\caption{Comparison of the dynamical evolutions of the expectation value of the dipole operator $D$ is illustrated for the exact (black line) and MF solutions (orange line with circles) and for SMF solutions with Gaussian (G) (red line with boxes), uniform (U) (green line with triangles), and two-point (T) distributions (blue line with diamonds) (a). The dynamics of the one-body entropy per particle $S/N$ for the exact and SMF solutions with the same three distributions is indicated in the lower panel (b). The interaction strength is $v_0=0.5\,\Delta$ and the initial state is $|\Psi(0)\rangle_1$.}
\label{fig5}
\end{figure}

The dynamics in the strong coupling case, $v_0=0.5\,\Delta$, is indicated in Fig.~\ref{fig5}. It is seen that the MF solution of $\langle D\rangle$ deviates from the exact one at $t=0.7\,\Delta^{-1}$ and SMF solutions start to deviate later at around $t=2\,\Delta^{-1}$. In Fig.~\ref{fig5}(a) and~\ref{fig5}(b), the difference between the SMF results with the three distributions is almost negligible. However, when the plots at $t=2.1\,\Delta^{-1}$ are zoomed in, as seen from the inset figures, the SMF solution with the two-point distribution turns out to be the best approximation to the exact solution and the SMF solution with the Gaussian distribution is the worst approximation. It is known that the validity time of MF as well as beyond-MF approximations decreases in inverse proportion to the coupling strength, $\Delta t_{\text{val}}\propto v_0^{-1}$,~\cite{lacroix2014a,polkovnikov2003}. Here, we observe the same behavior by comparing Figs.~\ref{fig3} and~\ref{fig5}. Furthermore, as the coupling strength increases the difference between SMF evolutions with different initial distributions decreases.

\begin{figure}[!t]
\vspace{0.2cm}
\includegraphics*[width=8.5cm]{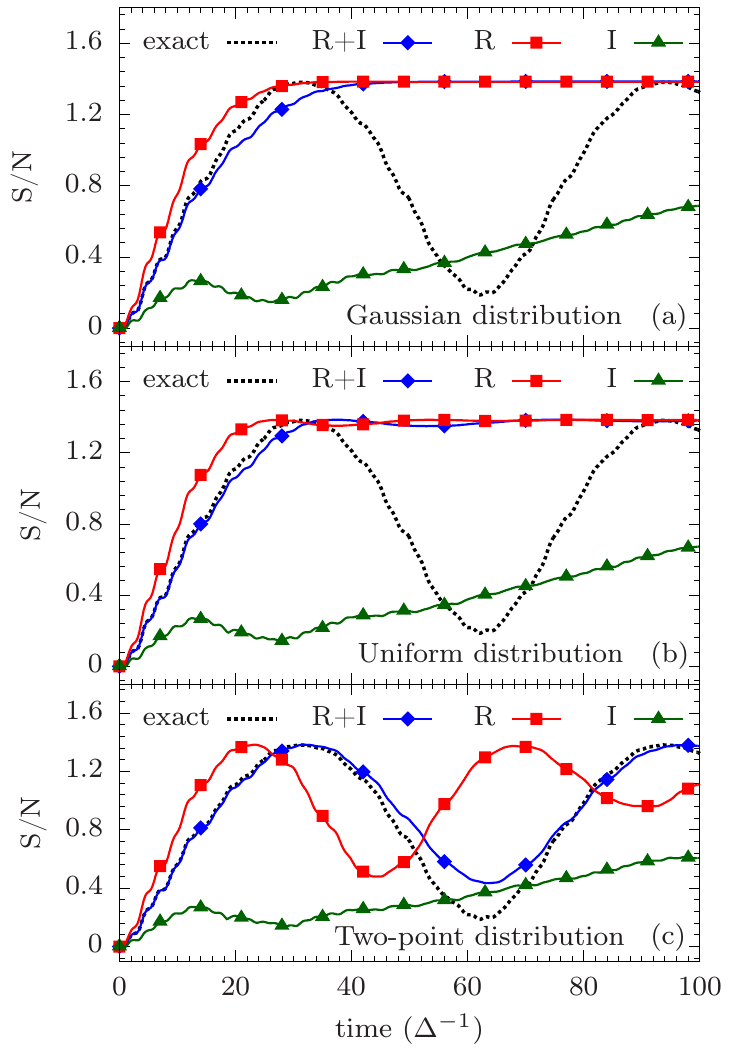}
\vspace{0.2cm}
\caption{SMF evolutions of the entropy per particle $S/N$ are plotted for different weights of the variances to the real $r_{ij}$ and imaginary $s_{ij}$ parts of the stochastic density matrix elements. The SMF solutions are obtained with the Gaussian (a), uniform (b), and two-point (c) distributions. The exact solutions are indicated by black lines. The SMF evolutions with two-point distribution are shown for: the equal weight case, $\overline{r^2_{ij}}=\overline{s^2_{ij}}=1/4$, which is labeled as R+I (blue line with diamonds); the full weight to the real parts, $\overline{r^2_{ij}}=1/2$ and $\overline{s^2_{ij}}=0$, which is labeled as R (red line with boxes); and the full weight to the imaginary parts, $\overline{s^2_{ij}}=1/2$ and $\overline{r^2_{ij}}=0$, which is labeled as I (green line with triangles). The interaction strength is $v_0=0.05\,\Delta$ and the initial state is $|\Psi(0)\rangle_1$.}
\label{fig6}
\end{figure}
\begin{figure}[!t]
\vspace{0.2cm}
\includegraphics*[width=8.5cm]{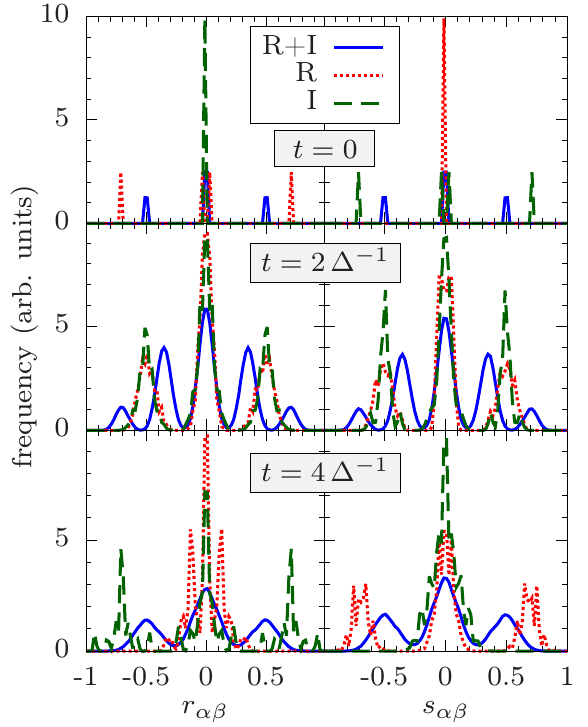}
\vspace{0.2cm}
\caption{Distributions of the real $r_{\alpha\beta}$ (the subfigures on the left) and imaginary $s_{\alpha\beta}$ parts (the subfigures on the right) of the matrix elements $\rho^\lambda_{\alpha\beta}$ of the stochastic one-body densities $\rho^\lambda$ at the times $t=0$ (upper subfigures), $t=2\,\Delta^{-1}$ (middle subfigures), and $t=4\,\Delta^{-1}$ (lower subfigures). The single particle states are $\alpha=(+1,+1/2)$ and $\beta=(-1,+3/2)$. The initial distribution of the matrix elements is the two-point distribution. The same cases as in Fig.~\ref{fig6} namely R+I (blue solid line), R (red dotted line), and I (green dashed line) are indicated. The interaction strength is $v_0=0.05\,\Delta$ and the initial state is $|\Psi(0)\rangle_1$.}
\label{fig7}
\end{figure}
\begin{figure}[!t]
\vspace{0.2cm}
\includegraphics*[width=8.3cm]{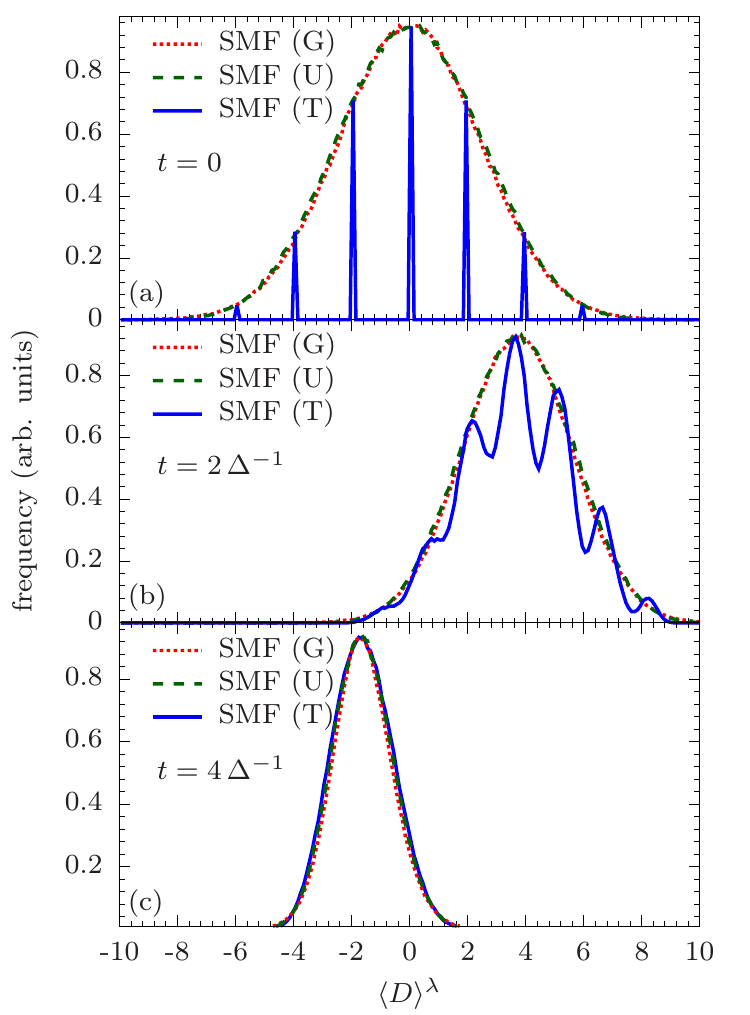}
\vspace{0.2cm}
\caption{Distribution of the ''event`` expectation values of the dipole operator $D$ in the ensemble are compared for the Gaussian (G) (red dotted line), uniform (U) (green dashed line), and two-point (T) distributions (blue solid line) at the times $t=0$ (a), $t=2\,\Delta^{-1}$ (b), $t=4\,\Delta^{-1}$ (c). The interaction strength is $v_0=0.05\,\Delta$ and the initial state is $|\Psi(0)\rangle_1$.}
\label{fig8}
\end{figure}
\begin{figure}[!t]
\vspace{0.2cm}
\includegraphics*[width=8.35cm]{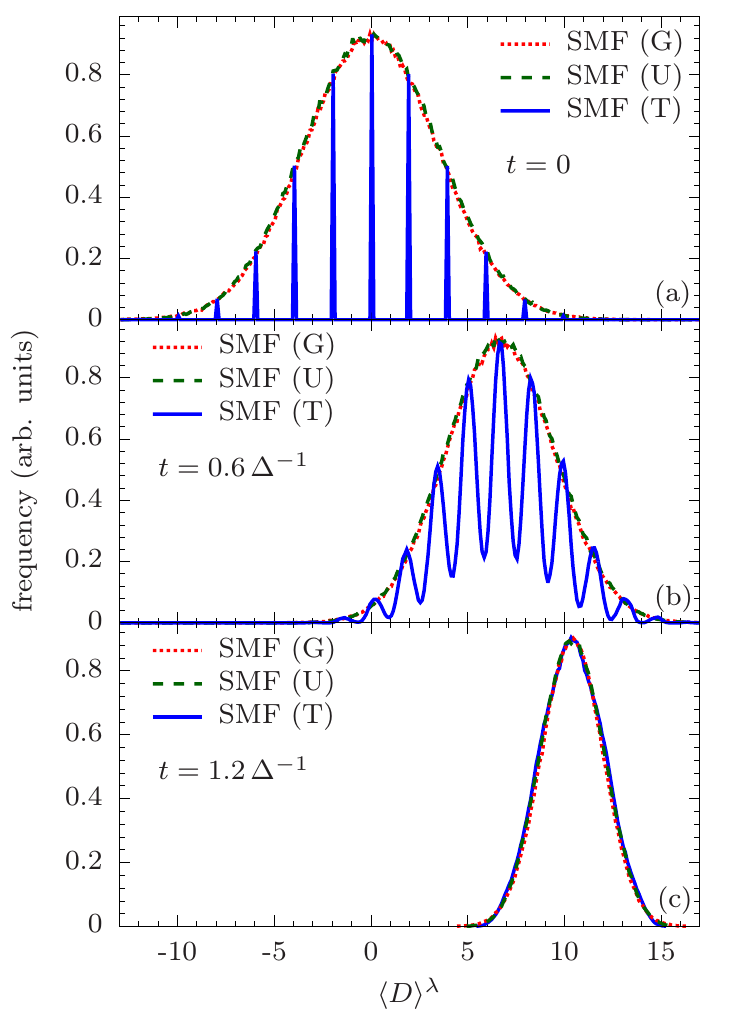}
\vspace{0.2cm}
\caption{Same as Fig.~\ref{fig8} except that the system consists of $N=12$ particles in 24 single particle states. The distributions of $\langle D\rangle^\lambda$ are compared at the times $t=0$ (a), $t=0.6\,\Delta^{-1}$ (b), $t=1.2\,\Delta^{-1}$ (c).}
\label{fig9}
\end{figure}
In Fig.~\ref{fig6}, we investigate the dependence of the SMF dynamics on the parameter $\chi$ that controls the weights of the real and imaginary parts of the stochastic matrix elements of the one-body densities $\rho^\lambda$. Three cases are considered: the first one corresponds to the equal weights case, labeled as R+I, for which $\chi=\overline{r^2_{ij}}=\overline{s^2_{ij}}=1/4$; the second one corresponds to full weight to the real parts, labeled as R, for which $\chi=\overline{r^2_{ij}}=1/2$ and $\overline{s^2_{ij}}=0$; and the third case is the opposite, full weight to the imaginary parts, labeled as I, for which $\overline{s^2_{ij}}=1/2$ and $\chi=\overline{r^2_{ij}}=0$. It is clearly seen that the equal weight case gives the best dynamics with all three distributions. When the imaginary matrix elements are set to zero, case R, the SMF entropy is overestimating the exact one and when the real matrix elements are set to zero, case I, the SMF entropy is underestimating the exact entropy. As discussed in Sec.~\ref{sec2}, the quality of the SMF approximation increases when the function $F$, given by Eq.~(\ref{func}), assumes smaller values. In Figs.~\ref{fig6}(a) and~\ref{fig6}(b), the SMF solutions with the Gaussian and uniform distributions are consistent with this result. However, the weight parameter $\chi$ should not affect the quality of the SMF evolution with the two-point distribution since $F$ is constant for this distribution, as seen from Fig.~\ref{fig1}. In Fig.~\ref{fig6}(c), it is observed that the equal weight case is still much better than unequal weight cases.
This can be explained by observing the distributions of real and imaginary parts of the stochastic matrix elements. As an illustration, in Fig.~\ref{fig7}, we show the distributions of the real $r_{\alpha\beta}$ and imaginary $s_{\alpha\beta}$ parts of a matrix element $\rho^\lambda_{\alpha\beta}$ of the stochastic one-body densities $\rho^\lambda$ at the times $t=0$, $t=2\,\Delta^{-1}$, and $t=4\,\Delta^{-1}$.  In this figure, the single particle states are arbitrarily chosen as $\alpha=(+1,+1/2)$ and $\beta=(-1,+3/2)$. The initial distribution of the matrix elements is the two-point distribution. The same cases as in Fig.~\ref{fig6} namely R+I, R, and I, are indicated. It is observed that when the imaginary (real) parts of the matrix elements are set to zero, case R (I), the real (imaginary) parts in the figure assume three values with equal probability whereas, the imaginary (real) parts are zero at the initial time $t=0$. After a very short time interval, at $t=2\,\Delta^{-1}$, both the real and imaginary parts develop similar three-peak structures. However, after the same amount of time, at $t=4\,\Delta^{-1}$, opposite distribution structures for the R and I cases are observed with respect to the distributions at time $t=0$. At time $t=4\,\Delta^{-1}$, the real (imaginary) parts develop a narrow distribution around the value zero compared to the imaginary (real) parts that have a wide three-peak distribution for the case R (I). The case R+I exhibits almost symmetrical distributions for the real and imaginary parts at all times. These results clearly show how the dynamical correlations between the real and imaginary parts of the matrix elements build up in time. In particular, the real and imaginary parts are balancing each other for the equal weight case R+I, whereas the distributions are changing between narrow single-peak around zero and a wide three-peak distribution for the unequal weights cases R and I. In Fig.~\ref{fig6}, it is seen that the entropies for the cases R and I start to deviate from the exact and R+I case solutions at very short times such as $t=2\,\Delta^{-1}$ for all the distributions. At this time, even the MF solution agrees well with the exact solution as seen from Fig.~\ref{fig3}(a). Based on these observations and the fact that the correlations between the real and imaginary parts of the matrix elements of the one-body density are governed by the MF equation, we think that the quality of the SMF approach for different weights cases, such as the cases R and I, is unrelated to the choice of the initial distributions in the SMF approach. Hence, the equal weight case R+I should always be considered as a better approximation over unequal cases due to dynamical correlations, governed by the MF equation, between the real and imaginary parts of the matrix elements.
\begin{figure}[!t]
\vspace{0.2cm}
\includegraphics*[width=8.5cm]{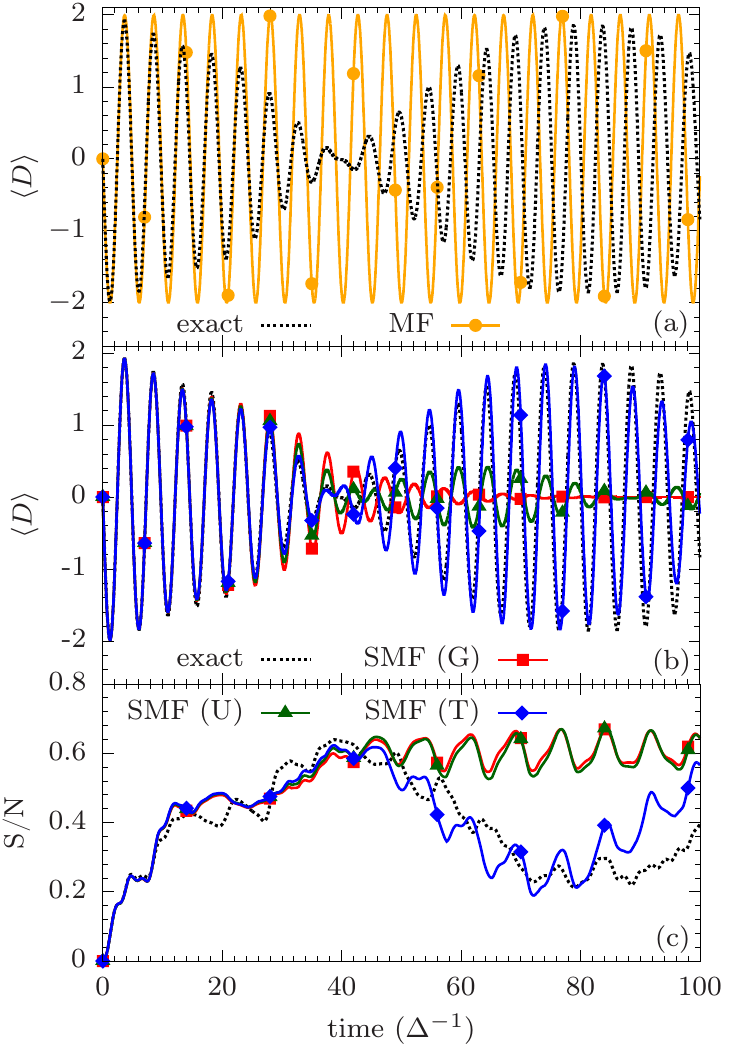}
\vspace{0.2cm}
\caption{Same as Fig.~\ref{fig3} except that the initial state is $|\Psi(0)\rangle_2$.}
\label{fig10}
\end{figure}
\begin{figure}[!t]
\vspace{0.2cm}
\includegraphics*[width=8.5cm]{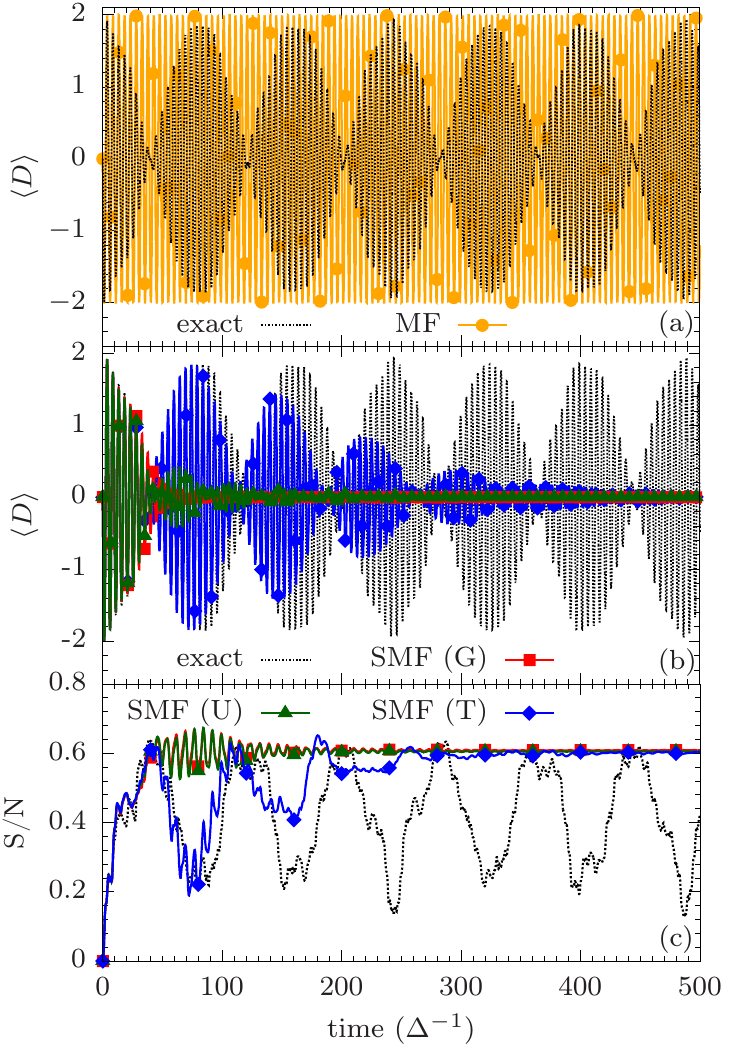}
\vspace{0.2cm}
\caption{Same as Fig.~\ref{fig10} for long time evolution}
\label{fig11}
\end{figure}

Figure~\ref{fig8} shows the distribution of the ''event`` expectation value, defined by Eq.~(\ref{ola}), of the collective observable $D$ for the three probability distributions at three different times. The distributions of $\langle D\rangle^\lambda$ are continuous and Gaussian for the initial Gaussian and uniform distributions at all times. However, for the initial two-point distribution, the resulting distribution of $\langle D\rangle^\lambda$ is discrete. This result is directly related to the fact that the two-point distribution is discrete. It is seen that in a very short time interval $4\,\Delta^{-1}$ the result with the two-point distribution becomes also Gaussian since the correlations between the stochastic matrix elements of the one-body density, governed by the stochastic version of the mean-field equation (\ref{lmgmf}), build up very fast. Figure~\ref{fig9} is the same as Fig.~\ref{fig8} except that we consider a system with $N=12$ particles in 24 single-particle states. Here, it is observed that the number of discete peaks increases linearly with particle number and also that the time interval after which a Gaussian distribution is obtained is even shorter than that for $N=6$ particles. The behaviors here are closely related to the central limit theorem, which states that the sum of many independent random variables tends toward a Gaussian distribution. Note that Figs.~\ref{fig8} and~\ref{fig9} show how fast the discrete initial distribution of $\langle D\rangle^\lambda$ transforms to a Gaussian distribution and hence how fast the correlations between the matrix elements of the one-body density build up. If the dynamical evolution of the distribution of $\langle D\rangle^\lambda$ in Figs.~\ref{fig8} or~\ref{fig9} is followed for longer times (not shown for brevity), it is observed that the Gaussian shape is preserved; however, centroids and widths of the distributions of $\langle D\rangle^\lambda$ for the two-point, Gaussian, and uniform distributions deviate from each other since the corresponding SMF means and variances of $D$ deviate from each other as well.

Similar results are reached for the SMF evolutions with different initial states. Here, we also demonstrate a result with a different initial state $|\Psi(0)\rangle_2$, which is explained in Sec.~\ref{subsec2}. The corresponding solutions are illustrated in Figs.~\ref{fig10} and~\ref{fig11}. Note that these figures are the same as Figs.~\ref{fig3} and~\ref{fig4} except that the initial state is different. The MF evolution with this initial state results in a sinusoidal oscillation with constant amplitude for the dipole moment $\langle D\rangle$, and the exact evolution demonstrates a beating oscillation with almost equal amplitudes for the beatings, as seen in Figs.~\ref{fig10}(a) and~\ref{fig11}(a). We see that taking a different initial state leads to the same conclusion that the SMF evolution with the two-point distribution leads to a better agreement with the exact evolution compared to the SMF evolution with the Gaussian or uniform distributions. In Figs.~\ref{fig10}(c) and~\ref{fig11}(c) the system does not reach the maximum entropy value $2\ln2$ due to the symmetry of the initial state. This is also in accordance with observation of the beating oscillations of the exact evolution. SMF evolution with the two-point distribution is able to reproduce the beatings with decreasing amplitudes in time whereas SMF evolutions with Gaussian and uniform distributions get damped in a very short time.  

\section{Conclusions}
\label{sec4}
In the SMF approach, expectation values of observables are obtained by statistically averaging over an ensemble of stochastic one-body densities. It is through matching these expectation values to the corresponding quantum expectation values of collective observables at initial time that provides information about the properties of the initial distribution of the matrix elements of the stochastic one-body densities. In the original formulation of the SMF approach, the matching of the expectation values is performed for the quantum means and variances of collective observables, which provides relations for the means and (co)variances (first and second moments) of the stochastic one-body density matrix elements that are then assumed to be Gaussian random numbers~\cite{ayik2008}. In the present work, we have investigated the properties of the stochastic one-body densities within the SMF approach by considering higher order moments. An expression for the fourth central moments of the matrix elements is derived which suggest that a probability distribution with minimal kurtosis for the initial density matrix elements provides the best approximation to the exact dynamics when the stochastic matrix elements are assumed to be statistically independent random numbers. The distribution with minimal kurtosis is the so-called two-point distribution, which can take only two values and hence is discrete. The SMF approach with the two-point distribution as well as the Gaussian and uniform distributions are applied to an exactly solvable model that is inspired by the Lipkin-Meshkov-Glick model that is familiar from nuclear physics~\cite{lacombe2016}.  It is shown that indeed the two-point distribution is the best approximation for the matrix elements of the stochastic one-body densities in the weak as well as strong coupling strengths. However, in the strong coupling case, the difference among the SMF evolutions with different distributions is almost negligible. 

MF dynamics in its various forms such as TDHF and density functional theory is widely used in many fields of physics. However, MF dynamics is known to contain some drawbacks. Quantum correlations in collective observables are mostly missing hampering the use of the MF dynamics for long time evolutions. Furthermore, the dissipation of the collective motion is severly underestimated. The SMF approach cures these drawbacks to some extent by taking into account the initial quantum fluctuations. It can be shown by following ref.~\cite{lacroix2016} that SMF is equivalent to solving a BBGKY-like hierarchy of equations describing the coupling between the different moments of the one-body density. As a consequence, the kth moment is coupled to the (k+1)th moment. The dissipation in one-body space emerges from the coupling with the second moment that is also coupled to higher order moments. This situation is rather similar to the case with the standard BBGKY hierarchy of equations. The equations of motion of the different moments are independent of the assumption for the initial probability. However, obviously, the quality of the result itself is anticipated to depend on how accurately not only the second moments but also higher order moments of the collective observables are described by the initial probability. This is actually what we show here: that a better account of moments higher than the second moment improves the SMF evolution. Hence, it is observed that the long time correlations are much better reproduced with the two-point distribution than the Gaussian or uniform distributions, which result in overdamping and hence faster thermalization. 

The discrete form of the two-point distribution results in a discrete initial distributions of the collective observables with the tips of the discrete values having a Gaussian shape. Due to correlations between the matrix elements supplied by the self-consistent SMF equation, the distribution of the collective observables reaches a Gaussian distribution very fast, which is in accordance with the central limit theorem. The discrete nature of the physical observables and the matrix elements of the initial stochastic one-body densities opens up the possibility to perform SMF dynamics with small numbers of events.         

\begin{acknowledgments}
B.Y. gratefully acknowledges useful discussions with Phuong Mai Dinh. This work was supported in part by the Scientific and Technological Research Council of Turkey (TUBITAK).
\end{acknowledgments}

\appendix*
\section{Quantum moments of a one-body operator}
Let us consider that a many-body system is described by a Slater determinant and that the corresponding one-body density is $\rho$. In terms of the natural basis $\{|i\rangle\}$, which satisfies $\langle i|\rho|j\rangle=n_i\delta_{ij}$ where the occupation numbers $n_i$ can take values 0 or 1, a one-body operator can be written as $A=\sum_{ij}A_{ij}a_i^\dag a_j$. The first moment is the expectation value given by
\begin{align}
\langle A\rangle=&{\rm Tr}({\rho} A)\nonumber\\
=&\sum_i n_i A_{ii}.
\end{align}
The second moment is obtained as
\begin{align}
\langle A^2\rangle
=&{\rm Tr_1}({\rho}_1 A^2_1) + {\rm Tr_{12}}({\rho}_{12} A_1A_2)\nonumber\\
=&\sum_{ij} n_i(1-n_j) A_{ij}A_{ji}+\sum_{ij} n_in_j A_{ii}A_{jj},
\end{align}
where $\rho_{12}=\rho_1\rho_2(1-P_{12})$. 
Hence, the second central moment is given by
\begin{align}
\langle\left(A-\langle A\rangle\right)^2\rangle
=&\langle A^2\rangle-\langle A\rangle^2\nonumber\\
=&\sum_{ij} n_i(1-n_j) A_{ji}A_{ij}\nonumber\\
=&\frac{1}{2}\sum_{ij} \left[n_i(1-n_j) + n_j(1-n_i)\right] A_{ji}A_{ij}.
\end{align}
The third moment of $A$ reads
\begin{align}
\langle A^3\rangle
=&{\rm Tr_1}({\rho}_1 A^3_1) + 2{\rm Tr_{12}}({\rho}_{12} A_1^2A_2)
+{\rm Tr_{12}}({\rho}_{12} A_1A_2^2)\nonumber\\
&+{\rm Tr_{123}}({\rho}_{123} A_1A_2A_3)\nonumber\\
=&\sum_{ijk} n_i A_{ij}A_{jk}A_{ki}\nonumber\\
&+2\sum_{ijk} n_i n_j \left(A_{ik}A_{ki}A_{jj}-A_{jk}A_{ki}A_{ij}\right)\nonumber\\
&+\sum_{ijk} n_i n_j \left(A_{ii}A_{jk}A_{kj}-A_{ji}A_{ik}A_{kj}\right)\nonumber\\
&+\sum_{ijk} n_i n_j n_k(A_{ii}A_{jj}A_{kk}-A_{ii}A_{jk}A_{kj}\nonumber\\
&\qquad\qquad\qquad-A_{ij}A_{ji}A_{kk}-A_{ik}A_{jj}A_{ki}\nonumber\\
&\qquad\qquad\qquad+A_{ij}A_{jk}A_{ki}+A_{ji}A_{kj}A_{ik})\nonumber\\
=&\sum_{ijk} n_i A_{ij}A_{jk}A_{ki}\nonumber\\
&+3\sum_{ijk} n_i n_j \left(A_{ik}A_{ki}A_{jj}-A_{jk}A_{ki}A_{ij}\right)\nonumber\\
&+\sum_{ijk} n_i n_j n_k(A_{ii}A_{jj}A_{kk}-3A_{ii}A_{jk}A_{kj}\nonumber\\
&\qquad\qquad\qquad+2A_{ij}A_{jk}A_{ki}),
\end{align}
where $\rho_{123}=\rho_1\rho_2\rho_3(1-P_{12})(1-P_{13}-P_{23})$. 
The third central moment of $A$ reads
\begin{align}
\langle\left(A-\langle A\rangle\right)^3\rangle
=&\langle A^3\rangle-3\langle A^2\rangle\langle A\rangle+2\langle A\rangle^3\nonumber\\
=&\sum_{ijk} n_i(1-3n_j+2n_jn_k) A_{ij}A_{jk}A_{ki}\nonumber\\
=&\sum_{ijk} n_i(1-3n_j)(1-n_jn_k) A_{ij}A_{jk}A_{ki}\nonumber\\
=&\sum_{ijk} \Lambda_{ijk}\, A_{ij}A_{jk}A_{ki},
\end{align}
where we used the fact that $n_i^2=n_i$ for a Slater determinant and $\Lambda_{ijk}$ is the symmetrized version of the term $n_i(1-3n_j)(1-n_jn_k)$ given by
\begin{align}
\label{p3A}
\Lambda^{(3)}_{ijk}=&\; \frac{1}{3}\left[n_i(1-3n_j)(1-n_jn_k)+n_k(1-3n_i)(1-n_in_j)\right.\nonumber\\
&\quad \left. + n_j(1-3n_k)(1-n_kn_i)\right].
\end{align}
The fourth moment of $A$ reads
\begin{align}
\langle A^4\rangle
=&{\rm Tr_1}({\rho}_1 A^4_1) + {\rm Tr_{12}}\left[{\rho}_{12} (4A_1^3A_2+3A_1^2A_2^2)\right]\nonumber\\
&+6{\rm Tr_{123}}({\rho}_{123} A_1^2A_2A_3)\nonumber\\
&+{\rm Tr_{1234}}({\rho}_{1234} A_1A_2A_3A_4)\nonumber\\
=&\sum_{ijkl} n_i A_{ij}A_{jk}A_{kl}A_{li}\nonumber\\
&+4\sum_{ijkl} n_i n_j \left(A_{ik}A_{kl}A_{li}A_{jj}-A_{jk}A_{kl}A_{li}A_{ij}\right)
\nonumber\\
&+3\sum_{ijkl} n_i n_j (A_{ik}A_{ki}A_{jl}A_{lj}-A_{jk}A_{ki}A_{il}A_{lj})\nonumber\\
&+6\sum_{ijkl} n_i n_j n_k(A_{il}A_{li}A_{jj}A_{kk}-A_{il}A_{li}A_{jk}A_{kj}\nonumber\\
&\qquad\qquad\qquad-A_{kl}A_{li}A_{jj}A_{ik}-A_{jl}A_{li}A_{ij}A_{kk}\nonumber\\
&\qquad\qquad\qquad+2A_{ij}A_{jk}A_{kl}A_{li})\nonumber\\
&+\sum_{ijkl}n_in_jn_kn_l(A_{ii}A_{jj}A_{kk}A_{ll}-6A_{ij}A_{ji}A_{kk}A_{ll}\nonumber\\
&\qquad\qquad\qquad+8A_{ij}A_{jk}A_{ki}A_{ll}+3A_{ij}A_{ji}A_{kl}A_{lk}\nonumber\\
&\qquad\qquad\qquad-6A_{ij}A_{jk}A_{kl}A_{li}),
\end{align}
where $\rho_{1234}=\rho_1\rho_2\rho_3\rho_4(1-P_{12})(1-P_{13}-P_{23})(1-P_{14}-P_{24}-P_{34})$. 
The fourth central moment of $A$ reads
\begin{align}
\langle\left(A-\langle A\rangle\right)^4\rangle
=&\langle A^4\rangle-4\langle A^3\rangle\langle A\rangle+6\langle A^2\rangle\langle A\rangle^2-3\langle A\rangle^4\nonumber\\
=&\sum_{ijkl} n_i(1-4n_j-3n_k+12n_jn_k-6n_jn_kn_l)\nonumber\\
&\quad\times A_{ij}A_{jk}A_{kl}A_{li}\nonumber\\
&+3\sum_{ijkl} n_in_k(1-2n_l+n_ln_j)\nonumber\\
&\quad\times A_{ij}A_{ji}A_{kl}A_{lk}\nonumber\\
=&\sum_{ijkl} n_i(1-4n_j)(1-3n_k)(1-n_jn_kn_l)\nonumber\\
&\quad\times A_{ij}A_{jk}A_{kl}A_{li}\nonumber\\
&+3\sum_{ijkl} n_in_k(1-2n_l)(1-n_ln_j)\nonumber\\
&\quad\times A_{ij}A_{ji}A_{kl}A_{lk}\label{sim}\\
=&\sum_{ijkl}\Lambda^{(4a)}_{ijkl}\,A_{ij}A_{jk}A_{kl}A_{li}\nonumber\\
&+3\sum_{ijkl}\Lambda^{(4b)}_{ijkl}\,A_{ij}A_{ji}A_{kl}A_{lk},
\end{align}
where the symmetrized terms are
\begin{align}
\Lambda^{(4a)}_{ijkl}=&\;\frac{1}{4}\left[ n_i(1-4n_j)(1-3n_k)(1-n_jn_kn_l)\right.\nonumber\\
&\quad+n_l(1-4n_i)(1-3n_j)(1-n_in_jn_k)\nonumber\\
&\quad+n_k(1-4n_l)(1-3n_i)(1-n_ln_in_j)\nonumber\\
&\quad\left.+n_j(1-4n_k)(1-3n_l)(1-n_kn_ln_i)\right],\\
\Lambda^{(4b)}_{ijkl} =&\;\frac{1}{8}\left\{ n_in_k\left[(1-2n_l)(1-n_ln_j)+(1-2n_j)(1-n_jn_l)\right]\right.\nonumber\\
&\quad+n_jn_k\left[(1-2n_l)(1-n_ln_i)+(1-2n_i)(1-n_in_l)\right]\nonumber\\
&\quad+n_in_l\left[(1-2n_k)(1-n_kn_j)+(1-2n_j)(1-n_jn_k)\right]\nonumber\\
&\quad\left.+n_jn_l\left[(1-2n_k)(1-n_kn_i)+(1-2n_i)(1-n_in_k)\right]\right\}.
\end{align}
Note that the symmetrized terms $\Lambda^{(4a)}_{ijkl}$ and $\Lambda^{(4b)}_{ijkl}$ are obtained by realizing, from Eq.~(\ref{sim}), the following equalities
\begin{align}
A_{ij}A_{jk}A_{kl}A_{li}&=A_{li}A_{ij}A_{jk}A_{kl}\nonumber\\
&=A_{kl}A_{li}A_{ij}A_{jk}\nonumber\\
&=A_{jk}A_{kl}A_{li}A_{ij}
\end{align}
and 
\begin{align}
A_{ij}A_{ji}A_{kl}A_{lk}&=A_{kl}A_{lk}A_{ij}A_{ji}\nonumber\\
&=A_{ji}A_{ij}A_{kl}A_{lk}\nonumber\\
&=A_{kl}A_{lk}A_{ji}A_{ij}\nonumber\\
&=A_{ij}A_{ji}A_{lk}A_{kl}\nonumber\\
&=A_{lk}A_{kl}A_{ij}A_{ji}\nonumber\\
&=A_{ji}A_{ij}A_{lk}A_{kl}\nonumber\\
&=A_{lk}A_{kl}A_{ji}A_{ij}.
\end{align}

\bibliography{VU_bibtex_master}

\end{document}